\documentclass[aps,prl,twocolumn, showpacs,amsmath,amssymb]{revtex4-1}
\usepackage{graphicx}
\usepackage{tensor}

\newcommand{\be}{\begin{equation}}
\newcommand{\ee}{\end{equation}}
\newcommand{\ba}{\begin{eqnarray}}
\newcommand{\ea}{\end{eqnarray}}

\newcommand{\nn}{\nonumber}

\begin{document}

\title{Superfluid Black Holes}

\author{Robie A. Hennigar}
\email{rhennigar@uwaterloo.ca}
\affiliation{Department of Physics and Astronomy, University of Waterloo,
Waterloo, Ontario, Canada, N2L 3G1}

\author{Robert B. Mann}
\email{rbmann@uwaterloo.ca}
\affiliation{Department of Physics and Astronomy, University of Waterloo,
Waterloo, Ontario, Canada, N2L 3G1}

\author{Erickson Tjoa}
\email{etjoa002@e.ntu.edu.sg}
\affiliation{Department of Physics and Astronomy, University of Waterloo,
Waterloo, Ontario, Canada, N2L 3G1}
\affiliation{Division of Physics and Applied Physics, Nanyang Technological University, 
Singapore, S637371}

\begin{abstract}
We present what we believe is the first example of a ``$\lambda$-line" phase transition in black hole thermodynamics.  This is a line of (continuous) second order phase transitions which in the case of liquid $^4$He marks the onset of superfluidity.  The phase transition occurs for a class of asymptotically AdS hairy black holes in Lovelock gravity where a real scalar field is conformally coupled to gravity. We discuss the origin of this phase transition and outline the circumstances under which it (or generalizations of it) could occur.
\end{abstract}


\maketitle

The study of black hole thermodynamics provides valuable insight into quantum properties of the gravitational field.  The thermodynamics of anti de Sitter black holes has been of great interest since the pioneering work of Hawking and Page which demonstrated the existence of a thermal radiation/large AdS black hole phase transition~\cite{Hawking:1982dh}.  Furthermore, these spacetimes admit a gauge duality description via a dual thermal field theory. 

Recently there has been interest in treating the cosmological constant as a thermodynamic variable \cite{CreightonMann:1995} which plays the role of pressure in the first law~\cite{Caldarelli:1999xj, Kastor:2009wy, Cvetic:2010jb}.  Within this context, the black hole mass takes on the interpretation of enthalpy and a number of connections with ordinary thermodynamics emerge.  It was shown that the thermodynamic behaviour of a charged AdS black hole is analogous to the van der Waals liquid/gas system, with the role of the liquid/gas transition played by a small/large black hole phase transition~\cite{Kubiznak:2012wp}.  Subsequent work has revealed examples of \textit{triple points}~\cite{Altamirano:2013uqa}, {\it (multiple) reentrant phase transitions}~\cite{Altamirano:2013ane, Frassino:2014pha}, \textit{isolated critical points}~\cite{Dolan:2014vba, Hennigar:2015esa} and a host of other behaviour  for black holes (see \cite{Kubiznak:2016qmn} and references therein for a review).

Here we present the first example of a line of second order (continuous) black hole phase transitions which strongly resemble those which occur in condensed matter systems, e.g. the onset of superfluidity in liquid helium~\cite{RevModPhys.71.S318}.  The phase transition  occurs in a broad class of asymptotically AdS hairy black holes in Lovelock gravity.  Lovelock gravity~\cite{Lovelock:1971yv} is a geometric higher curvature theory of gravity and is the natural generalization of Einstein gravity to higher dimensions, giving rise to second order field equations for the metric.   It provides an excellent testbed for examining the effects of higher curvature corrections which appear in, for example, the low energy effective action of string theory~\cite{Zwiebach:1985uq}.  Recently, it has been shown \cite{Oliva:2011np} that a scalar field can be conformally coupled to the Lovelock terms and the resulting theory gives rise to analytic hairy black hole solutions \cite{Giribet:2014bva} evading no-go results which had been reported previously~\cite{nogo_hairy}.  These solutions have already been shown to possess interesting thermodynamic properties~\cite{Giribet:2014fla, Galante:2015voa, Hennigar:2015wxa} (such as reentrant phase transitions), and are of inherent interest due to the role scalar hair plays in holography, e.g. in descriptions of holographic superconductors and superfluids~\cite{Hartnoll:2008vx, Nie:2015zia}.

The model we consider consists of Lovelock gravity, a Maxwell field, and a real scalar field coupled conformally to the dimensionally extended Euler densities,
\be\label{action} 
{\cal I} =  \frac{1}{16 \pi G}\int d^dx \sqrt{-g} \, \left(\sum_{k=0}^{k_{\rm max}} {\cal L}^{(k)} - 4 \pi G F_{\mu\nu}F^{\mu\nu} \right)
\ee
where 
\begin{align}
{\cal  L}^{(k)} = &  \frac{1}{2^k} \delta^{(k)} \left(a_k \prod_r^k R^{{\alpha_{r}}{\beta_{r}}}_{{\mu_r} {\nu_r}} \right. \left. + b_k \phi^{d-4k}  \prod_r^k S^{{\alpha_{r}}{\beta_{r}}}_{{\mu_r} {\nu_r}}  \right)
\end{align}
with $\delta^{(k)} = \delta^{\alpha_1 \beta_1 \cdots \alpha_k \beta_k}_{\mu_1 \nu_1 \cdots \mu_k \nu_k}$ the generalized Kronecker tensor.  Here the tensor $\tensor{S}{_\mu_\nu^\gamma^\delta}$  describes how the scalar field couples to gravity,
\begin{align} 
\tensor{S}{_\mu_\nu^\gamma^\delta} =& \phi^2 \tensor{R}{_\mu_\nu^\gamma^\delta} - 2 \delta^{[\gamma}_{[\mu} \delta^{\delta]}_{\nu]}\nabla_\rho\phi\nabla^\rho\phi 
\nn\\
&- 4\phi \delta^{[\gamma}_{[\mu} \nabla_{\nu]} \nabla^{\delta]} \phi + 8 \delta^{[\gamma}_{[\mu}\nabla_{\nu]}\phi\nabla^{\delta]}\phi \,,
\end{align}
and transforms homogeneously under the conformal transformation, $g_{\mu\nu} \to \Omega^2 g_{\mu\nu}$ and $\phi \to \Omega^{-1}\phi$ as $\tensor{S}{_\mu_\nu^\gamma^\delta} \to \Omega^{-4} \tensor{S}{_\mu_\nu^\gamma^\delta}$.

We take a line element of the form,
\be 
ds^2 = -f dt^2 + f^{-1} dr^2 + r^2 d \Sigma^2_{(\sigma)d-2}
\ee
where $d \Sigma^2_{(\sigma)d-2}$ is the line element on a surface of constant curvature $\sigma (d-2)(d-3)$ with $\sigma = +1, 0, -1$ corresponding to spherical, flat and hyperbolic geometries; in the latter cases, the space is compact via identification~\cite{Mann:1997iz}. For this ansatz, the field equations for the metric reduce to,
\begin{align}\label{mastereqn} 
\sum_{k=0}^{k_{\rm max}} \alpha_k \left(\frac{\sigma-f}{r^2} \right)^k =& \frac{16 \pi G M}{(d-2)\Sigma^{\sigma}_{d-2} r^{d-1}} \nn\\
&+ \frac{H}{r^d} - \frac{8 \pi G}{(d-2)(d-3)}\frac{Q^2}{ r^{2d-4}} \, 
\end{align}
where
\begin{align}
\alpha_0 &= \frac{a_0}{(d-1)(d-2)}\, , \quad \alpha_1 = a_1 \, , \nn\\
 \alpha_k &= a_k \prod_{n=3}^{2k} (d-n) \, \textrm{ for } k \ge 2\, ,
\end{align}
and 
\be\label{defn_of_H} 
H = \sum_{k=0}^{k_{\rm max}} \frac{(d-3)!}{(d-2(k+1))!}b_k \sigma^k N^{d-2k} \, 
\ee
is the ``hair parameter".  For this configuration the scalar field takes the form,
\be 
\phi = \frac{N}{r}
\ee
and its equations of motion reduce to the following constraints:
\ba 
\sum_{k=1}^{k_{\rm max}} k b_k \frac{(d-1)!}{(d-2k-1)!} \sigma^{k-1}N^{2-2k} &=& 0 \, ,
\nn\\
\sum_{k=0}^{k_{\rm max}} b_k \frac{(d-1)! \left(d(d-1)+4k^2 \right)}{(d-2k-1)!} \sigma^k N^{-2k} &=& 0 \, .
\label{N-constr}
\ea
Since these are two equations in a single unknown ($N$), one equation enforces a constraint on the allowed coupling constants, $b_k$.  Computing the temperature by requiring the absence of conical singularities in the Euclidean sector and the entropy using the Iyer-Wald formalism~\cite{Iyer:1994ys}, we find the thermodynamic quantities for this solution are
\begin{widetext}
\begin{align}
\label{temp_etc} 
M &= \frac{(d-2) \Sigma_{d-2}^{\sigma}}{16 \pi G} \sum_{k=0}^{k_{\rm max}} \alpha_k \sigma^k r_+^{d - 2k -1} - \frac{(d-2)\Sigma^{\sigma}_{d-2} H}{16 \pi G r_+} + \frac{\Sigma^{\sigma}_{d-2} Q^2}{2(d-3) r_+^{d-3}}
\nn\\
T &= \frac{f'(r_+)}{4 \pi} = \frac{1}{4 \pi r_+ D(r_+)} \left[\sum_k \sigma \alpha_k(d-2k-1) \left(\frac{\sigma}{r_+^2}\right)^{k-1} + \frac{H}{r_+^{d-2}} - \frac{8 \pi G Q^2}{(d-2) r_+^{2(d-3)}} \right]
\nn\\
S &= \frac{\Sigma^{(\sigma)}_{d-2}}{4 G}   \left[ \sum_{k=1}^{k_{\rm max}} \frac{(d-2) k \sigma^{k-1}  \alpha_k  }{d-2k} r_+^{d-2k}  - \frac{d}{2\sigma (d-4)} H\right] \, \quad \textrm{if } b_k = 0 \,\,\,  \forall k > 2.
\end{align}
\end{widetext}
where $D(r_+) = \sum_{k=1}^{k_{\rm max}} k \alpha_k (\sigma r_+^{-2})^{k-1} $.  It is straightforward to show that
they satisfy the extended first law and Smarr formula.  As the remaining expressions are quite lengthy  we shall report them elsewhere~\cite{HennigarHair}.

In what follows we consider  $\alpha_k = 0 \,  \forall k > 3$ and $b_k = 0 \, \forall k > 2$.  This last condition is for simplicity: the falloff in the metric function is the same for all $b_k$ and the contribution to the entropy is always just a constant; so only the first three $b_k$'s are required to see all the physics of the scalar hair.  

Introducing the dimensionless parameters
\begin{align}
r_+ &= v\alpha_3^{1/4} \, , \quad  T = \frac{t\alpha_3^{-1/4}}{d-2} \, , \quad  H =\frac{4\pi h}{d-2}\alpha_3^{\frac{d-2}{4}}\nn\\
Q &= \frac{q}{\sqrt{2}}\alpha_3^{\frac{d-3}{4}} \, , \quad  m = \frac{16\pi M}{(d-2)\Sigma_{d-2}^{(\kappa)}\alpha_3^{\frac{d-3}{4}}} \nn\\
p&=\frac{\alpha_0(d-1)(d-2)\sqrt{\alpha_3}}{4\pi} \, , \quad \alpha=\frac{\alpha_2}{\sqrt{\alpha_3}}\, ,
\nn\\
G &=   M- TS  = \alpha_3^{-\frac{(d-3)}{4}} g \, .
\end{align}
The dimensionless equation of state (obtained by solving the expression for the temperature in Eq.~\eqref{temp_etc} for the pressure) reads
\begin{align}\label{eos}
p = &\frac{t}{v}-\frac{\sigma(d-3)(d-2)}{4\pi v^2}+\frac{2\alpha\sigma t}{v^3}-\frac{\alpha(d-2)(d-5)}{4\pi v^4}+ \frac{3t}{v^5}\nn\\&-\frac{\sigma(d-7)(d-2)}{4\pi v^6}+\frac{q^2}{v^{2(d-2)}}-\frac{h}{v^d}
\end{align}
where the quantity $p$ represents the pressure and $g$  the dimensionless Gibbs free energy.  At equilibrium, the state of the system is that which globally minimizes the Gibbs free energy.  

\begin{figure*}[!htp]
\centering
\includegraphics[scale=0.3]{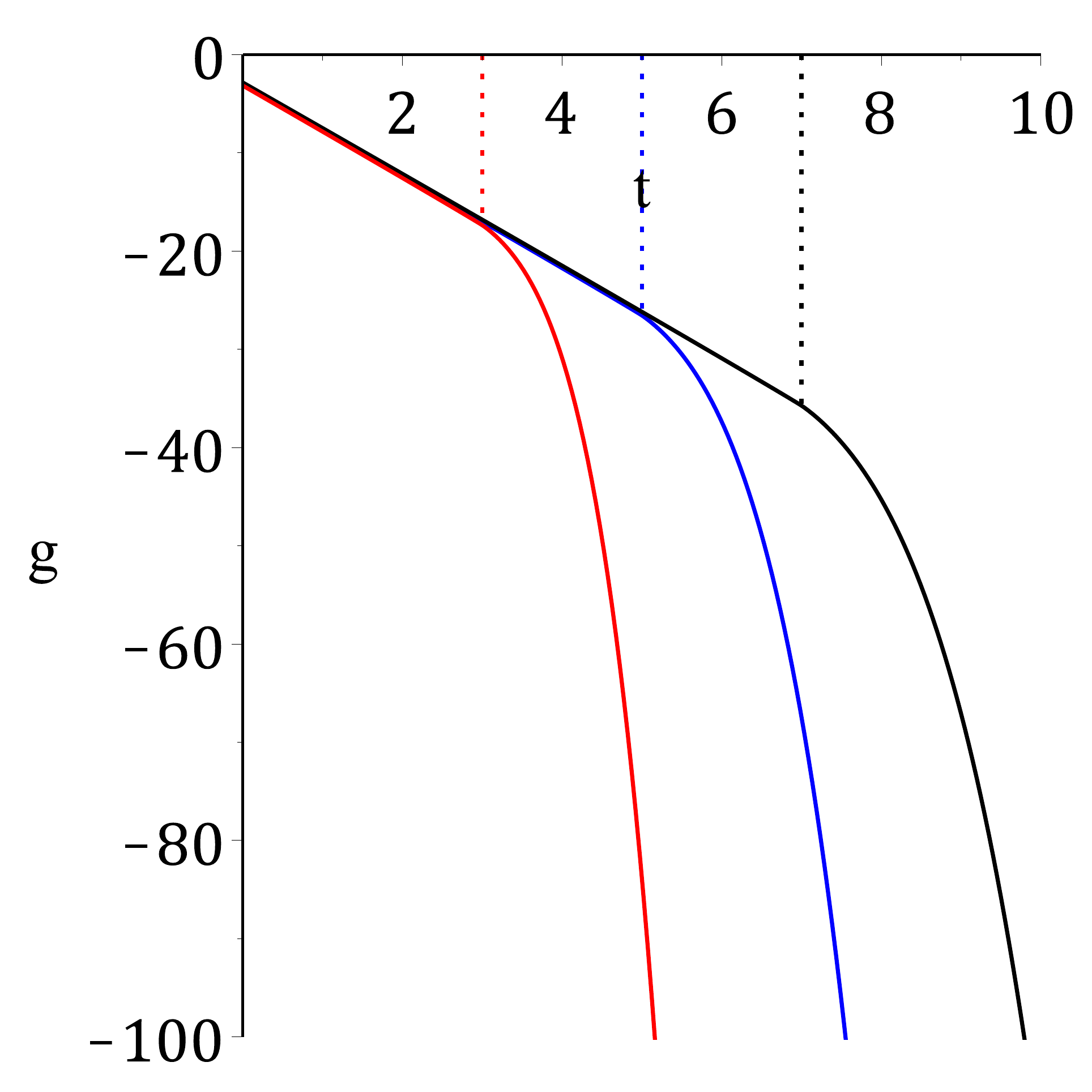}
\includegraphics[scale=0.3]{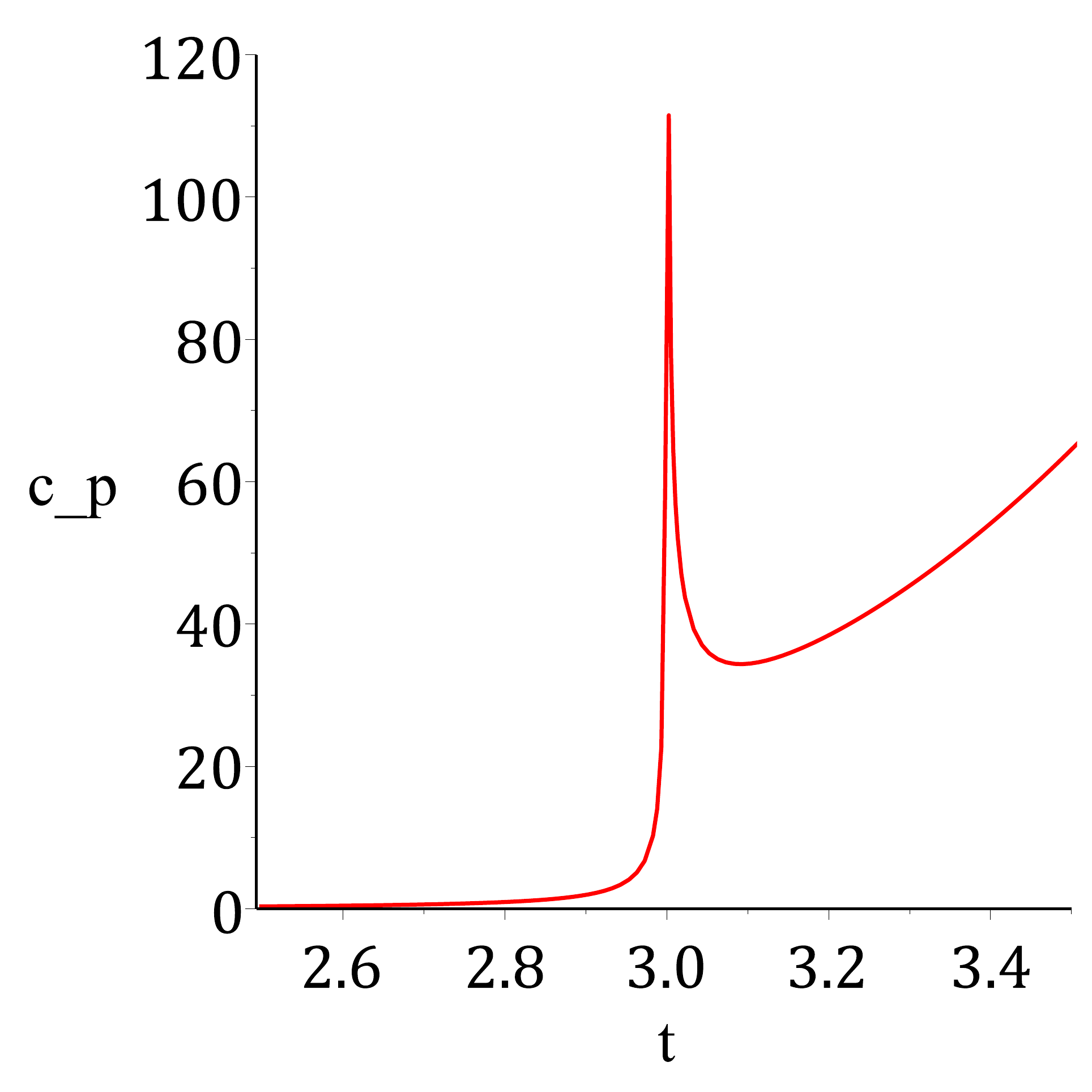}
\includegraphics[scale=0.3]{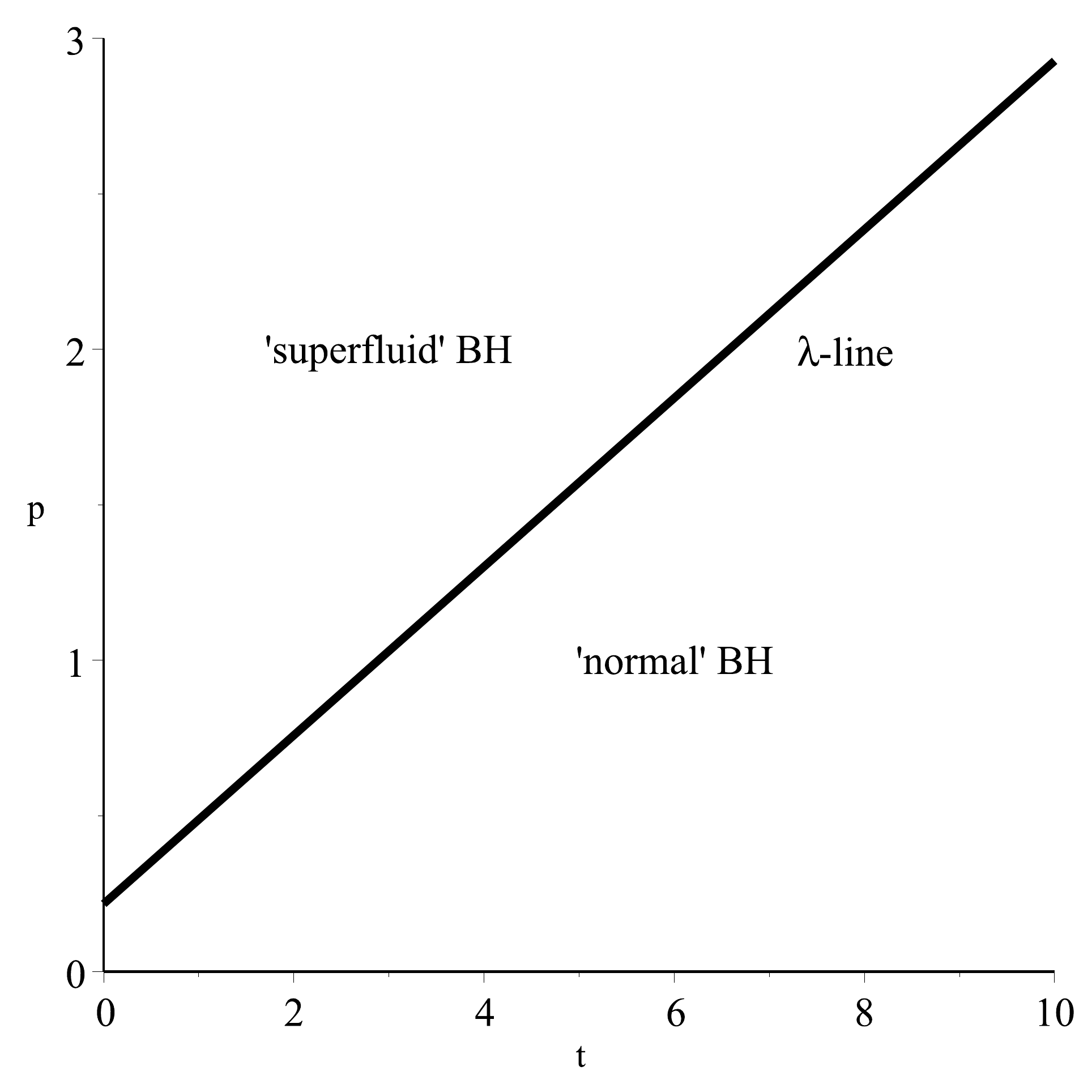}
\caption{{\bf Thermodynamic behaviour near $\lambda$ transition}: \textit{Left}: A plot of the Gibbs free energy vs. temperature for three distinct pressures chosen so that critical temperatures are  $t_c=3,5,7$ corresponding to the red, blue and black curves.  The dotted lines highlight the points where the second derivative of the Gibbs free energy diverges.  \textit{Center}: A plot of the specific heat $c_p=-t\frac{\partial^2 g}{\partial t^2}$ for the case $t_c = 3$.  \textit{Right}: $p$--$t$ parameter space.  The black line shows the locus of critical points, i.e. a line of second-order phase transitions known as the `lambda' line in the context of superfluidity.  These plots are for $d=7$.}
\label{cl_diverginggibbs}
\end{figure*}

Noting that the conditions for a critical point are
\be\label{eqn:cp_condition} 
\frac{\partial p}{\partial v}=\frac{\partial^2 p}{\partial v^2}=0  
\ee 
we find that for $\alpha = \sqrt{5/3}$ if $h$ and $q$ are set to the values
\be 
 h=\frac{4(2d-5)(d-2)^2v_c^{d-6}}{\pi d (d-4)}, \, \, q^2= \frac{2(d-1)(d-2) v_c^{2d-10}}{ \pi (d-4)}
\ee
and $\sigma = -1$,  Eq.~\eqref{eqn:cp_condition} is satisfied by $v_c = 15^{1/4}$ and 
\be 
p_c  = \left[ \frac{8}{225} (15)^{\frac{3}{4}} \right]t_c + \frac{\sqrt{15} (11d-40)(d-1)(d-2)}{900 \pi d}
\ee
\textit{for all temperatures} $t_c$!  In other words, this system exhibits \textit{infinitely many critical points} with critical volume $v_c = 15^{1/4}$.  In the $p-v$ plane, every isotherm is a critical isotherm, i.e. has an inflection point at $v= 15^{1/4}$.  In the variables $(t, p)$ there is no first order phase transition but rather a line of second order phase transitions, characterized by a diverging specific heat $c_p = -t \, \partial ^2 g / \partial t^2$ at the critical values.  We show representative thermodynamic behaviour in Fig.~\ref{cl_diverginggibbs} for $d=7$.  

\begin{figure}[!htp]
\centering
\includegraphics[width=0.3\textwidth]{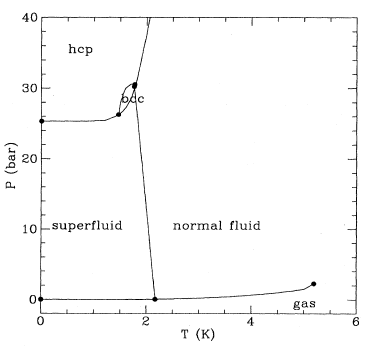}
\includegraphics[width=0.3\textwidth]{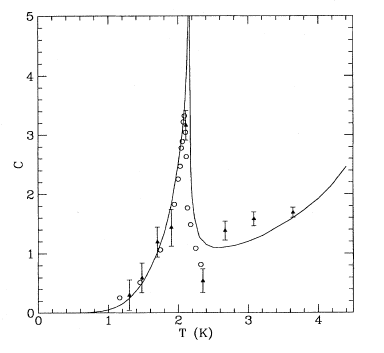}
\caption{{\bf Thermodynamic properties of $^4$He}: \textit{Top}: The $P$--$T$ phase diagram for $^4$He. The $\lambda$ line corresponds to a line of critical points where a second order phase transition occurs marking the onset of \textit{superfluidity}.  To the left of the $\lambda$-line, the liquid begins to exhibit remarkable properties. \textit{Bottom}: The specific heat of liquid $^4$He.  As the $\lambda$-line is approached, the specific heat spikes taking the shape of the Greek letter ``$\lambda$".  These plots have been reprinted with permission from~\cite{RevModPhys.67.279}.}
\label{fig:helium_plots}
\end{figure}

The line of second order phase transitions mimics those that occur in condensed matter systems where they correspond to, for example, fluid/superfluid transitions~\cite{RevModPhys.73.1}, superconductivity~\cite{superconductor}, and paramagentism/ferromagnetism transitions~\cite{Pathria2011401}. 	Building on the black hole/van der Waals fluid analogy~\cite{Kubiznak:2012wp}, the natural interpretation here is that this second order phase transition between small/large black holes corresponds to a fluid/superfluid type transition.   The resemblance to the fluid/superfluid $\lambda$-line transition of $^4$He (Fig.~\ref{fig:helium_plots}) is striking.  In each case, a line of critical points separates the two phases of `fluid' where specific heat takes on the same qualitative ``$\lambda"$ structure.  The phase diagram for helium is more complicated, including solid and gaseous states.  This is to be expected since helium is a complicated system, while these hairy black hole solutions are comparatively simple being characterized by only four numbers: $v, h, q$ and $\alpha$. However, it is remarkable that with so few parameters the essence of the $\lambda$-line can be captured. Most of the interesting properties of a superfluid are either dynamical or require a full quantum description to understand (see, e.g. \cite{basicSuperfluids, RevModPhys.71.S318} for an introduction and review).  Since we do not have access to a model of the underlying quantum degrees of freedom it is not possible to explore the black hole analogues of these properties at a deeper level. 

It is natural to wonder if there are any pathological properties of these black holes.  We have examined the Kretchmann scalar near the $\lambda$-line and have found it to be finite at all finite temperatures and pressures.  We have also studied the explicit solution to the field equations in detail and have found it to be completely regular outside the horizon.  Within the horizon and at large enough pressures, the first derivative of the metric function diverges.  Such behaviour is neither fatal nor uncommon for Lovelock black holes---similar behaviour occurs for charged black holes in Gauss-Bonnet gravity (cf. Fig.~2 of~\cite{Frassino:2014pha}).  The Gibbs free energy and temperature are continuous and differentiable near the critical point and the entropy is positive. The specific heat is positive, indicating thermodynamic stability of these black holes.  Furthermore, we have examined the linearized equations of motion for the theory about a maximally symmetric background and found that for the values of the coupling constants taken here the theory is free from ghost and tachyon instabilities~\cite{HennigarHair}.  Thus it seems that there is no underlying pathological behaviour here.  

To calculate valid critical exponents, the appropriate ordering field, $\Theta$, must be identified.   As for liquid helium~\cite{RevModPhys.73.1}, pressure is no longer the appropriate ordering field for this line of 2nd order phase transitions.   In this case there are three options for $\Theta$: $q$, $h$ or $\alpha$.  The resultant critical exponents are independent of which choice is made, but $q$ is the most natural choice since its 
variation does not entail any coupling constants.
Rearranging the expression for the temperature \eqref{temp_etc} for the chosen ordering field $\Theta$, it can be expanded near a critical point (in terms of $\omega = (v-v_c)/v_c$ and $\tau = (t- t_c)/t_c$) to give: 
\be \label{critexp}
\frac{\Theta}{\Theta_c} = 1 - A \tau + B \tau \omega   - C \omega^3  + {\cal O}(\tau \omega^2, \omega^4) \, ,
\ee 
where $A$, $B$ and $C$ are non-zero constants whose numeric value depends on the pressure and choice of ordering field.  It is clear from \eqref{critexp} that the critical exponents are
\be 
\alpha = 0 \, , \quad \beta = \frac{1}{2} \, \quad \gamma = 1 \, , \quad \delta = 3
\ee
and respectively govern the behaviour of the specific heat at constant volume, $C_V \propto |\tau|^{-\alpha}$, the order parameter $\omega \propto |\tau|^{\beta}$, the susceptibility/compressibility $(\partial \omega /\partial \Theta)|_{\tau}~\propto~|\tau|^{- \gamma}$ and the ordering field $\Theta \propto |\omega|^{\delta}$ near a critical point.  These results coincide with the mean field theory values, in particular for those for a superfluid in  $d > 5$ (cf. Table I of \cite{RevModPhys.73.1}).

We conclude by examining under what situations these $\lambda$-lines can be expected for black holes.  Here the key result was that the conditions for a critical point are satisfied without fixing the temperature.  For an equation of state of the form,
\be 
P = a_1(r_+, \varphi_i) \,  T + a_2(r_+, \varphi_i)
\ee
(where $\varphi_i$ represent additional constants in the equation of state) this condition is satisfied provided a nontrivial solution for the following equations exists:
\begin{align}\label{supercriteqs}
\frac{\partial a_i}{\partial r_+} &= 0 \, , \quad  \frac{\partial^2 a_i}{\partial r_+^2} = 0 \,  \quad i=1,2  \, .
\end{align}
With four free parameters ($v, \alpha, q$ and $h$), the hairy black holes permit a non-trivial solution to these 4 equations.  This result generalizes to all Lovelock theories cubic and higher: with an appropriate choice of parameters, they possess $\lambda$-lines in the presence of conformal hair and charge. However the necessary and sufficient conditions for satisfying \eqref{supercriteqs} for black holes
in general remain to be found. We have checked that neither the rotating black hole of $5d$ minimal gauged super-gravity~\cite{Chong:2005hr} nor those in higher order Lovelock gravity (without hair) admit a non-trivial solution.  A particularly interesting case would result if the above equations admitted two (or more) non-trivial solutions.  Such a circumstance could give rise to two intersecting $\lambda$-lines, a situation that occurs in certain ferromagnetic materials~\cite{0038-5670-24-1-R04}.

To summarize, we have presented the first example of a superfluid-like phase transition in black hole thermodynamics.  This occurs for hyperbolic black holes with conformal scalar hair in  cubic Lovelock gravity  for any dimension $d \ge 7$.  We have examined the black hole solutions and verified they are free from pathological behaviour.  We have presented precise conditions  \eqref{supercriteqs} that a black hole equation of state must satisfy to display similar behaviour.  While satisfaction of these conditions is by no means trivial, we find that
all cubic-and-higher Lovelock theories with conformal hair can satisfy these requirements. We suspect there exists a  much broader class of gravitational theories containing black holes that will exhibit $\lambda$-line phase transitions.  
 Determining further examples of this, or similar, behaviour remains an interesting problem for future research.

\section*{Acknowledgments}
This work was supported in part by the Natural Sciences and Engineering Research Council of Canada.  
  
\bibliographystyle{JHEP}
\bibliography{LBIB2}  
\end{document}